\documentclass[
    ,final            
  ,numberedheadings 
  ]
  {aipproc}

\layoutstyle{6x9}

\usepackage{amsmath,amsfonts,amssymb}
\def\nmh{{\sc nmhdecay}}
\def\asusy{a^{\rm SUSY}_\mu}
\def\relic{\Omega_{\neut}\,h^2}
\def\neut{\tilde\chi_1^0}
\def\neumass{m_{\tilde\chi_1^0}}
\newcommand{\crosssec}{\sigma_{\tilde\chi^0_1-p}}

\begin{document}

\title{NMSSM neutralino dark matter}



\author{Ana M. Teixeira}{address={Laboratoire de Physique Th\'eorique, 
UMR 8627,  Universit\'e de Paris-Sud XI,  B\^atiment 201, 
F-91405 Orsay Cedex, France }}
 
\begin{abstract}
We study the viability of the lightest neutralino as a
dark matter candidate in the Next-to-Minimal Supersymmetric
Standard Model. Taking into account accelerator
constraints as well as bounds on low-energy observables, and 
imposing consistency with present bounds on the 
neutralino relic density, we address the prospects for 
the direct detection of neutralino dark matter. We find regions of 
the allowed parameter space where the neutralino detection cross
section is within the reach of dark matter detectors, essentially owing 
to the presence of very light singlet-like Higgses, and to either singlino 
dominated or very light neutralinos.
\end{abstract}

\maketitle


The Next-to-Minimal Supersymmetric Standard Model (NMSSM) 
is a well-motivated extension of the Minimal Supersymmetric Standard
Model (MSSM) by a singlet superfield $\hat S$. In addition to
providing an elegant solution to the 
so-called $\mu$ problem of the MSSM, and rendering
the Higgs ``little fine tuning problem'' of the MSSM less severe, the 
presence of additional fields (CP-even, CP-odd neutral Higgs bosons,
and a fifth neutralino) leads to a richer and more complex 
phenomenology. In the NMSSM, one has possibility of dark matter 
scenarios that can be very different from those
encountered in the MSSM, both regarding the relic density as well as
prospects for direct detection. In particular, the exchange of very
light Higgses can lead to large direct detection cross sections,
within the reach of the present generation of dark matter
detectors~\cite{Cerdeno:2004xw,Cerdeno:2007sn}.

The NMSSM superpotential contains new couplings, involving the 
Higgs doublets and the singlet, 
\begin{equation}\label{Wnmssm:def}
  W_\mathrm{NMSSM}=
  \epsilon_{ij} \left(
  Y_u \, \hat H_2^j\, \hat Q^i \, \hat u +
  Y_d \, \hat H_1^i\, \hat Q^j \, \hat d +
  Y_e \, \hat H_1^i\, \hat L^j \, \hat e \right)
  - \epsilon_{ij} \lambda \,\hat S \,\hat H_1^i \hat H_2^j
  +\frac{1}{3} \kappa \hat S^3\,. 
\end{equation}
New terms associated with additional soft supersymmetry (SUSY) 
breaking trilinear couplings
$A_\lambda$ and $A_\kappa$ will also appear in the Lagrangian.
After the spontaneous breaking of electroweak (EW) symmetry, the neutral
Higgs scalars develop vacuum expectation values (VEVs), 
$\langle H_1^0 \rangle = v_1$, $\langle H_2^0 \rangle = v_2$ and 
$\langle S \rangle = s$. This leads to the dynamical generation
of an effective interaction $\mu \hat H_1 \hat H_2$, with 
$\mu \equiv \lambda s$.
In the NMSSM spectrum, we now have three CP-even and two CP-odd Higgs
states. In particular, the lightest Higgs scalar can be written as 
$h_1^0=S_{11} H^0_1 +S_{12} H^0_2 + S_{13} S$, where $S$ is the
unitary matrix that diagonalises the $3\times 3$ 
scalar Higgs mass matrix. In the
neutralino sector, the singlino mixes with the bino, wino and
Higgsinos. The lightest state can be now expressed as
$\tilde \chi^0_1 = N_{11} \tilde B^0 + N_{12} \tilde W_3^0 +
  N_{13} \tilde H_1^0 + N_{14} \tilde H_2^0 + N_{15} \tilde S$, where
  $N$ diagonalises the $5\times 5$ neutralino mass matrix.

The low-energy NMSSM parameter space can be described in terms of the
$\lambda, \, \kappa,\, \tan \beta,\, \mu,\, A_\lambda,$ $A_\kappa$ 
degrees of freedom,
as well as the soft SUSY-breaking terms, namely gaugino masses,
$M_{1,2,3}$. A thorough analysis of the low-energy 
NMSSM phenomenology (minimisation of the potential, computation of
spectrum and compatibility with LEP/Tevatron bounds) can be
obtained using the \nmh~2.0
code~\cite{Ellwanger:2005dv}. Additionally, we have also included in
our analysis~\cite{Cerdeno:2007sn} a more precise
computation of the $b\to s \gamma$ decay in the NMSSM~\cite{hiller},
taking into account next-to-leading order contributions, and imposing 
consistency at the $2\sigma$ level 
with the experimental central value~\cite{bsg_exp}, 
$  {\rm BR}^{\rm exp}(b\to s \gamma)=(3.55 \pm 0.27)\times10^{-4}\,.$
Likewise, we have also incorporated the constraints coming from the
contribution of a light pseudoscalar $a^0$ in the NMSSM to the rare $B$-
and $K$-meson decays~\cite{hiller}. Finally, in our analysis we have
also included the
constraints coming from the SUSY contributions to the muon anomalous
magnetic moment, $a_{\mu}=(g_{\mu}-2)$.
At present, the observed excess in
$a^{\rm exp}_{\mu}$~\cite{g-2_exp} constrains a possible
SUSY contribution to be~\cite{g2:sm}
$a_{\mu}^{\mathrm{SUSY}}=(27.6\,\pm\,8)\times 10^{-10}$.

As thoroughly discussed 
in~\cite{Cerdeno:2004xw,Cerdeno:2007sn}, an extensive part
of the low-energy NMSSM parameter space is directly excluded on
theoretical grounds, namely the occurrence of tachyons. Moreover, false
minima, and Landau poles for the couplings in $W_\mathrm{NMSSM}$ also 
render unviable further areas. In regions surviving the latter
constraints and the LEP/Tevatron bounds, and that present the most 
appealing prospects regarding dark matter direct detection, an
important role is played by the $b \to s\,\gamma$ decay, which can in
principle exclude important regions of the parameter space (due to
very large contributions to BR($b \to s\,\gamma$) from charged Higgs 
exchange). Concerning the SUSY contributions to $a_{\mu}$, in the 
dark-matter interesting regions, these tend to be 
in general quite small. A sufficiently large
$\asusy$ can nevertheless be obtained when slepton (and
gaugino) masses are decreased, in association with large values of the
slepton trilinear couplings.

In order to be a good dark matter candidate, the lightest
NMSSM neutralino must also comply with the increasingly stringent
bounds on its relic density, 
\begin{equation}\label{om:mi}
  0.1 \lesssim \Omega h^2 \lesssim 0.3\, \,\,\mathrm{(astrophysical)},
\quad \quad 
  0.095 \lesssim \Omega h^2 \lesssim 0.112\,\,\,\mathrm{(WMAP)},
 \end{equation}
respectively arising from astrophysical
 constraints~\cite{lightreview}, and from 
taking into account the recent three years data from the WMAP
satellite~\cite{wmap}. 

Compared to the MSSM, there are several alterations leading to $\relic$:
first, the possibility of a singlino-like lightest supersymmetric particle
(LSP), associated with new couplings in the interaction
Lagrangian, may favour the coupling of WIMPs to a singlet-like
Higgs, whose mass can be substantially lighter than in the MSSM, 
given the more relaxed experimental constraints. Secondly, 
in the NMSSM we have new open channels for neutralino annihilation,
e.g. $s$-channel resonances, new channels with annihilation  
into $Z\,h^0_1$, $h^0_1\,h^0_1$, $h^0_1\,a^0_1$ and $a^0_1\,a^0_1$
(due to light $h^0_1$ and $a_1^0$ states), providing important contributions
to the annihilation and co-annihilation 
cross-sections~\cite{Belanger:2005kh}. In our analysis~\cite{Cerdeno:2007sn},
the results for the neutralino relic density, obtained from an \nmh \,
link to MicrOMEGAS~\cite{Belanger:2005kh}.

In the regions of the parameter space likely to have large neutralino
detection cross sections~\cite{Cerdeno:2004xw}, we have found that, in
general, the correct relic density can only be obtained 
when either the singlino composition of the neutralino
is large enough or when the annihilation channels into $Z$, $W$, or
$h_1^0$ are kinematically forbidden. 
Interestingly, some allowed areas were very close to the
tachyonic border, which as we verified, can give rise to very large 
direct detection cross sections.

As pointed out in~\cite{Cerdeno:2004xw}, the existence of a fifth
neutralino state, together with the presence 
of new terms in the Higgs-neutralino-neutralino interaction
(which are proportional to $\lambda$ and $\kappa$), trigger new 
contributions to the spin-independent part of the
neutralino-nucleon cross section, $\sigma_{\tilde \chi^0_1 -p}$. 
On the one hand, although the
term associated with the $s$-channel squark exchange is formally
identical to the MSSM case, it can be significantly reduced if the
lightest neutralino has a major singlino composition. 
On the other hand, and more importantly, the dominant contribution to
$\sigma_{\tilde \chi^0_1 -p}$, associated to the exchange of CP-even
Higgs bosons on the $t$-channel, can be largely enhanced when these are
very light. Consequently, large detection cross sections can be obtained, 
even within the reach of the present generation of dark matter detectors.

\begin{center}
\begin{figure}[!t]\hspace*{0mm} 
\includegraphics[height=.33\textheight]{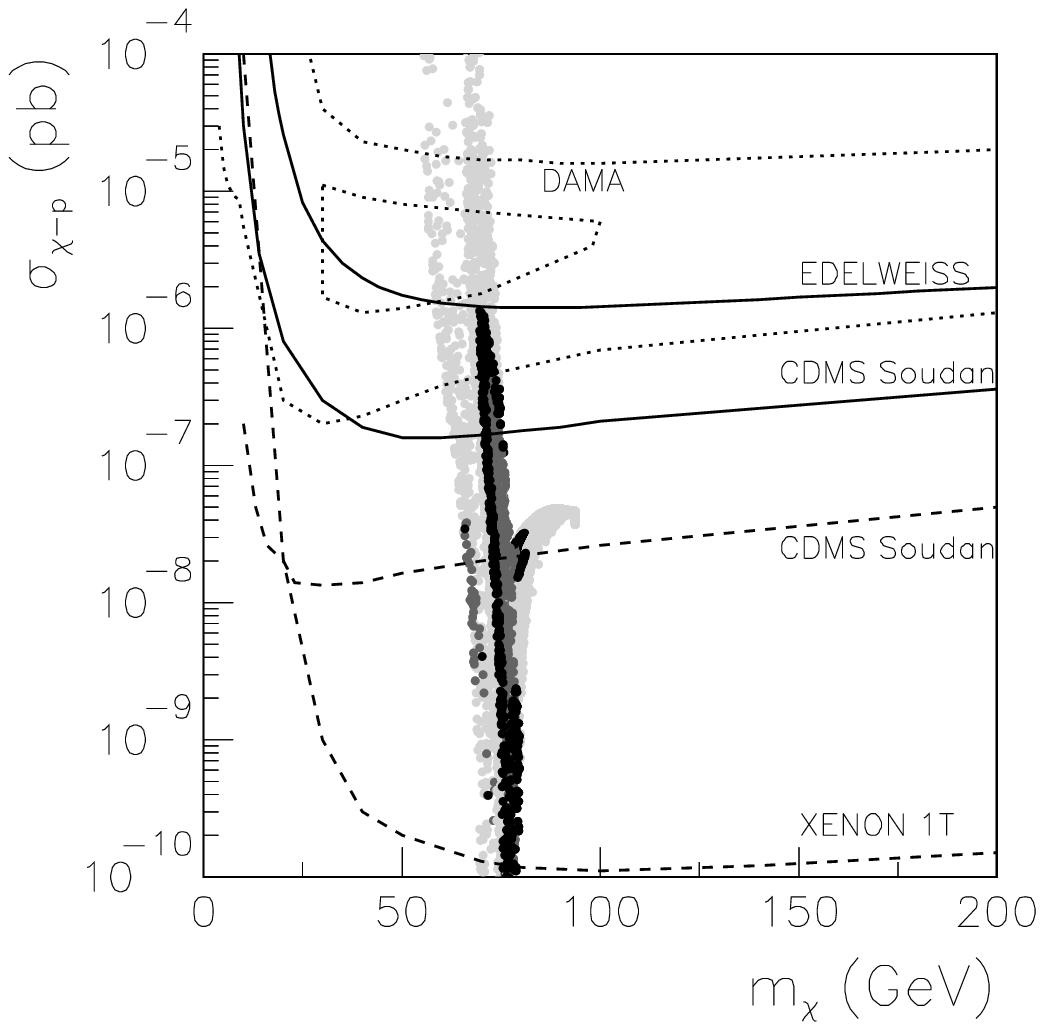}\hspace*{0mm} 
\includegraphics[height=.33\textheight]{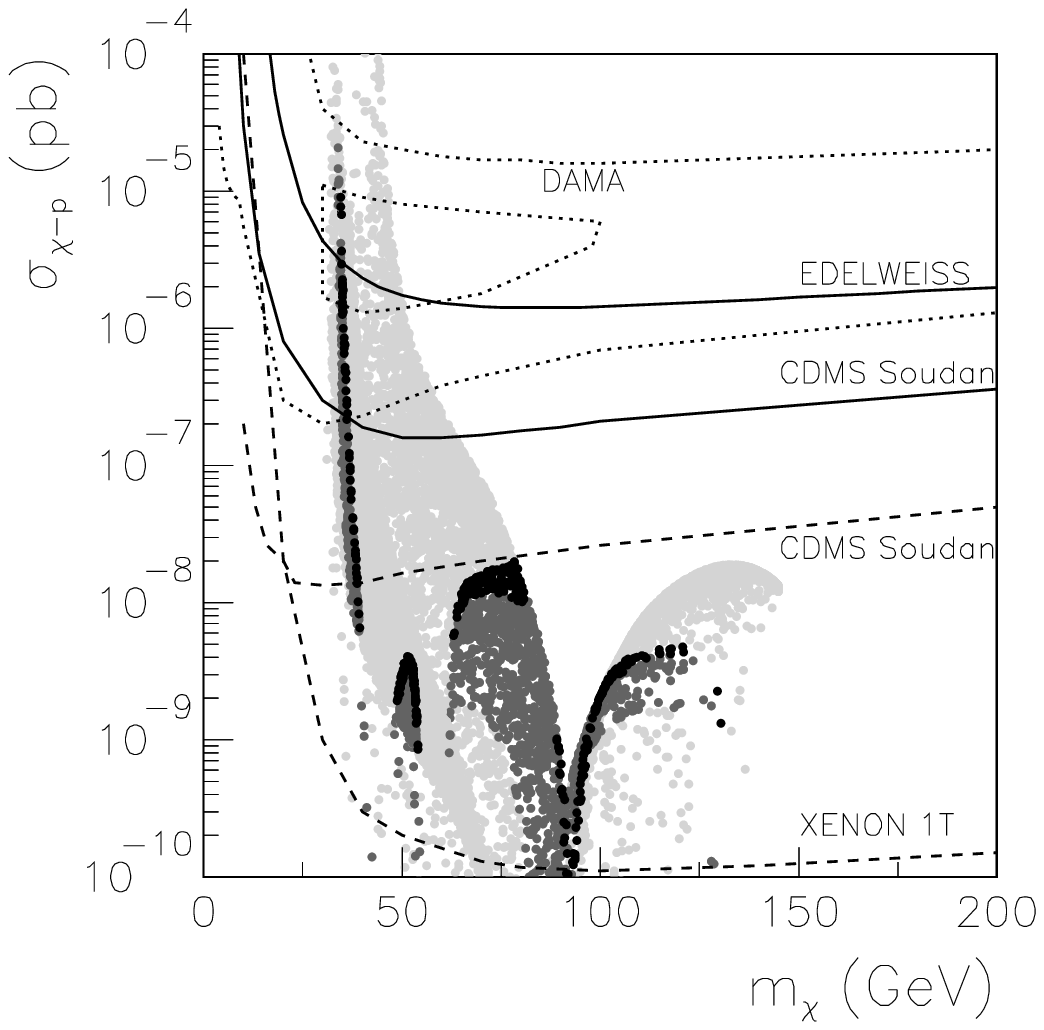}
  \vspace*{-.5cm}
  \caption{ 
    Scatter plot of the scalar neutralino-nucleon cross section as a
    function of the lightest neutralino mass. On the left, $M_1= 160$ 
    GeV, $\tan \beta=5$, $A_\lambda=400$~GeV, $A_\kappa=-200$ GeV, 
    and $\mu=130$ GeV.  All the points represented are in
    agreement with LEP/Tevatron, $\asusy$, and 
    BR($b\to s\,\gamma$) bounds. Dark grey dots
    represent points which, in addition, fulfil $0.1\le\relic\le0.3$,
    whereas black dots are those in agreement with the WMAP constraint. 
    The sensitivities of present and projected experiments are also 
    depicted, with solid and dashed lines, respectively. 
    On the right, a different example with $M_1= 330$ GeV, $\tan
    \beta =5$, $A_\lambda=570$   GeV, $A_\kappa=-60$ GeV, with
    $\mu=160$ GeV, a case where the resulting
    $\asusy$ is outside the experimental $2\sigma$ region.
  }
  \label{fig:svhn:160}
\end{figure}
\end{center}

As illustrative examples of our analysis, let us consider two cases,
which represent the most relevant features of NMSSM dark matter scenarios. 
Let us begin by considering $M_1=160$~GeV,
 $A_\lambda=400$~GeV, $A_\kappa=-200$~GeV, and  $\mu=130$~GeV, with
 $\tan \beta=5$, leading  to a parameter space consistent 
with bounds on $\asusy$ and BR($b \to s\,\gamma$).
In regions of the parameter space where the neutralino is relatively heavy
and has a mixed bino-Higgsino composition, 
the relic density is too small. 
As the neutralino mass decreases, some annihilation channels become
kinematically forbidden, such as annihilation into a
pair of $Z$ or $W$ bosons when $\neumass<M_Z$ or $\neumass<M_W$,
respectively. Below these, the resulting relic density can be 
large enough to fulfil the WMAP constraint, which occurs for distinct
regions in parameter space~\cite{Cerdeno:2007sn}. 
One of the allowed regions is close to the tachyonic area and 
exhibits very light singlet-like Higgses, potentially leading to large
detection cross sections.
This is indeed the case, as evidenced on the left-hand side of 
Fig.\,\ref{fig:svhn:160}, where the theoretical
predictions for $\crosssec$ are plotted versus the lightest 
neutralino mass. The resulting $\crosssec$ spans several orders of
magnitude, but, remarkably, areas with $\crosssec \gtrsim 10^{-7}$ pb 
are found. These correspond to the above mentioned regions of 
the parameter space with very light 
singlet-like Higgses ($25\,{\rm GeV}\,\lesssim m_{h_1^0}\lesssim 
50\,{\rm GeV}$ with $S_{13}^2\gtrsim0.99$). 
The neutralino is a mixed singlino-Higgsino state
($N_{15}^2\approx0.35$) with mass around $75$ GeV. The sensitivities of
present and projected dark matter experiments are also depicted 
for comparison.
On the right-hand side of Fig.\,\ref{fig:svhn:160} we show the
resulting $\crosssec$ when the neutralino composition is changed,
namely when the Higgsino component is enhanced. In particular, we 
consider the choice $M_1= 330$ GeV, $\tan \beta =5$, 
$A_\lambda=570$ GeV, $A_\kappa=-60$ GeV, with $\mu=160$ GeV.
Such neutralinos annihilate more efficiently, thus
leading to a reduced $\relic$, so that the astrophysical constraint
becomes more stringent. On the right-hand side of
Fig.\,\ref{fig:svhn:160}, the various resonances appear as funnels in the
predicted $\crosssec$ for the regions
with the correct $\relic$ at the corresponding values of the 
neutralino mass ($\neumass\approx M_Z/2$ and $\neumass\approx
m_{h_1^0}/2$). 
Below the resonance with the $Z$ boson, light neutralinos are obtained
$\neumass\lesssim M_Z/2$ with a large singlino composition which have
the correct relic abundance. The lightest Higgs is also singlet-like
and very light, leading to a very large detection cross section,
$\crosssec\gtrsim10^{-6}$ pb.

\begin{theacknowledgments}
The work of A.~M.~Teixeira was supported by the French ANR
project PHYS@COL\&COS. The work reported here was based in
collaborations with D.~G.~Cerde\~no, E.~Gabrielli, C.~Hugonie,
D.~E.~L\'opez-Fogliani and C.~Mu\~noz. 
\end{theacknowledgments}

\end{document}